\documentclass[aps,linenumbers,twocolumn,floatfix,superscriptaddress]{revtex4}
\usepackage{hyperref,color}
\usepackage{natbib}
\setcitestyle{square, comma, numbers,sort&compress, super}

\RequirePackage{graphicx,amsmath}
\RequirePackage{mathptmx}      
\RequirePackage{flushend}
\RequirePackage{lineno}

\usepackage{doi}
\usepackage[]{units}
\usepackage{xspace}
\usepackage{pbox}
\usepackage{multirow}
\usepackage{lineno}

\newcommand{\expp}{\ensuremath{e}} 
\newcommand{\I}{\ensuremath{\textrm{I}}} 
\newcommand{\Q}{\ensuremath{\textrm{Q}}} 
\newcommand{\power}{\ensuremath{\textrm{P}_{\mu w}}} 
\newcommand{\tph}{\ensuremath{\tau_\textrm{ph}}} 
\newcommand{\tphSim}{\ensuremath{T_\textrm{ph}^{10-90}}} 
\newcommand{\tqp}{\ensuremath{\tau_\textrm{qp}}} 
\newcommand{\TSiAl}{\ensuremath{\textrm{T}_{\textrm{Si}\rightarrow \textrm{Al}}}} 
\newcommand{\TSiTef}{\ensuremath{\textrm{T}_{\textrm{Si}\rightarrow \textrm{Tef}}}} 
\newcommand{\nuDebye}{\ensuremath{\nu_\textrm{Debye}}} 
\newcommand{\etapb}{\ensuremath{\eta_\textrm{pb}}} 
\newcommand{\etageom}{\ensuremath{\eta_{_{\textrm{KID}}}}} 
\newcommand{\Eabs}{\ensuremath{E_\textrm{abs}}} 
\newcommand{\nmax}{\ensuremath{n_\textrm{refl}^\textrm{max}}} 
\newcommand{\Tex}{\ensuremath{T_\textrm{ex}}} 
\newcommand{\tring}{\ensuremath{\tau_\textrm{ring}}} 
\newcommand{\Tqp}{\ensuremath{T_\textrm{qp}}} 
\newcommand{\Nqp}{\ensuremath{N_\textrm{qp}}} 
\newcommand{\Phiqp}{\ensuremath{\Phi_\textrm{qp}}} 

\newcommand{\DBD}{\ensuremath{0\nu DBD}}

\begin{document}

\title{Measurements and simulations of athermal phonon transmission from silicon absorbers to aluminium sensors}

\author{M. Martinez}
\email[]{mariam@unizar.es}
\affiliation{Laboratorio de F\'isica Nuclear y Astropart\'iculas, Universidad de Zaragoza, C/ Pedro Cerbuna 12, 50009 Zaragoza, Spain}
\affiliation{Fundaci\'on ARAID, Av. de Ranillas 1D,  50018 Zaragoza, Spain}
\author{L. Cardani}
\author{N. Casali}
\author{A. Cruciani}
\affiliation{INFN-Sezione di Roma, Piazzale Aldo Moro 2, 00185 Roma, Italy}
\author{G. Pettinari}
\affiliation{Istituto di Fotonica e Nanotecnologie-CNR, Via Cineto Romano 42, 00156 Roma, Italy}
\author{M. Vignati}
\affiliation{INFN-Sezione di Roma, Piazzale Aldo Moro 2, 00185 Roma, Italy}

\begin{abstract}
Phonon reflection/transmission at the interfaces plays a fundamental role in 
cryogenic particle detectors, in which the optimization of the phonon signal at the sensor 
(in case of phonon-mediated detectors) or the minimization of the heat transmission 
(when the detection occurs in the sensor itself) is of primary importance to improve sensitivity.
Nevertheless the mechanisms governing the phonon physics at the interfaces are still not completely understood.
The two more successful models, Acoustic Mismatch Model (AMM) and Diffuse Mismatch Model (DMM) are not able
to explain all the accumulated experimental data, and the measurement of the transmission coefficients between the materials remains a challenge. 
Here, we use measurements of the athermal phonon flux in aluminium Kinetic Inductance Detectors (KID) deposited
on silicon substrates following a particle interaction 
to validate a Monte Carlo (MC) phonon simulation. 
We apply the Mattis-Bardeen theory to derive the phonon pulse energy and timing from the KID signal and
compare the results with the MC for specular (AMM) and diffuse (DMM) reflection, finding a remarkable good 
agreement for specular, while diffuse reflection is clearly disfavoured. 
For an aluminum film of 60 nm and a silicon substrate of 380~$\mu$m, we obtain 
transmission coefficients Si-Al in the range [0.3~-~0.55] and Si-Teflon in the range [0.1~-~0.15]. 

\end{abstract}

\maketitle
\vline



\section{Introduction}
\label{sec:intro}
\indent\indent Phonon-mediated cryogenic detectors using massive absorbers 
are a mature technology extensively employed in rare event physics experiments,
like neutrinoless double beta decay ({\DBD}) searches~\cite{DellOro:2016tmg} (see for example CUORE~\cite{2018PhRvL.120m2501A},
CUPID-0~\cite{2018PhRvL.120w2502A}, LUMINEU~\cite{2015JInst..10P5007A}, AMoRE~\cite{2016ITNS...63..543L}...)
and dark matter direct detection experiments~\cite{Goodman:1984dc,Strigari:2013iaa} (EDELWEISS~\cite{2016EPJC...76..548H}
, SuperCDMS~\cite{dur24168}, CRESST~\cite{2016EPJC...76...25A}...). 
Generally the working temperature is $<$100~mK and the most common phonon sensors are 
Neutron Transmutation Doped (NTD) Ge or ion-implanted Si thermistors~\cite{McCammon}, 
which are glued or bonded to the detector surface 
and depending on the gluing characteristics are more or less sensitive to the ballistic component, 
or Transition Edge Sensors (TES)~\cite{Irwin}, which usually are sensitive to the ballistic phonons. 
Recently other kinds of sensor are being developed, as Metallic Magnetic Calorimeters (MMC)~\cite{MCC} 
or Kinetic Inductance Detectors (KIDs, the ones used in this work)~\cite{2003Natur.425..817D}. 
In all cases, low-threshold detection and/or an identification of the event topology 
(multi-site event, bulk vs surface...) and nature of the interacting particle ($\alpha$, $\beta/\gamma$, nuclear recoil...)
are mandatory, and hence a good understanding of the
phonon transport mechanism and heat losses in the interfaces is fundamental. 
On the other hand, a good understanding of these processes
could also be useful to mitigate the effects of unwanted 
phonon-mediated signals. This is the case of cryogenic bolometers employed for CMB measurements in space, 
which are severely affected by cosmic rays~\cite{2014A&A...569A..88C,2016A&A...592A..26C},
and superconducting qubits, where phonons generated by cosmic rays and natural radioactivity can modify the qubit state~\cite{2018PhRvL.121o7701S,2018PhRvL.121k7001G}. 
\par
Monte Carlo (MC) simulation of particles transport and interactions in matter is one of the basic ingredients 
for the design of a particle detector and the detection efficiency calculation. 
In particular the GEANT4 package~\cite{geant4}, initially developed for high energy Physics,
now is used by a much wider community, including astroparticle, space and medical Physics.
Nevertheless, at the level of phonon physics, despite the fundamental role they play in the energy
collection in cryogenic bolometers, there is no generalized use of this analysis tool.
Recently GEANT4 has incorporated 
condensed matter physics elements as phonon and electron-hole pairs,
essential for a more complete understanding of a cryogenic detector.
The code was first developed by the CDMS cryogenic Dark Matter experiment~\cite{Agnese:2015ywx} 
and subsequently integrated into GEANT4 as a general open-source package 
called G4CMP (GEANT4 Condensed Matter Physics)~\cite{Brandt:2014imy,g4cmp}. 
It has been validated for germanium, reproducing quite accurately the results of some
experiments using heat pulses (produced for example by a focused laser beam ) to excite  ballistic phonons~\cite{Brandt:2012zzb},
and also giving a good description of the CDMS detectors: 
Ge cylinders with interleaved ionization and grounded phonon electrodes coupled to  tungsten TES to read the phonon signal.
The MC simulation reproduces the arrival time of the ballistic phonons into the TES and 
the energy partition between the phonon and charge~\cite{Leman:2011cc,McCarthy:2011sx}.
\par
A correct treatment of the phonon scattering/transmission at the interfaces is 
a main ingredient of the simulation when the sensitive area is a small fraction of the total detector surface.
Nevertheless, phonon scattering at the interfaces is still an open question, and there is no general agreement about the
model to describe the
experimental data, being the most well established ones the acoustic mismatch model (AMM) and
the diffuse mismatch model (DMM)~\cite{1989RvMP...61..605S}.
AMM, which proposes specular reflection on the
interface in analogy with the Snell's law for light, has been very successful
at low temperatures~\cite{1987ApPhL..51.2200S}, while DMM, in which phonons undergo diffuse reflection, is sensitive to surface roughness
and preferred at temperatures above 1~K~\cite{2017PhRvB..95t5423H}.
G4CMP include a basic implementation of the phonon reflection mechanism based on these models 
in which a phonon in the boundary undergoes a reflection (specular for the AMM model or  diffuse (Lambertian) for the DMM model), or is transmitted through the boundary with a certain probability given by a transmission coefficient,
but the experimental determination of the
phonon transmission coefficients at the interfaces is a hard task, and currently large uncertainties exist.
\par
In this work we apply the G4CMP package to model two prototypes of the CALDER project~\cite{Battistelli:2015vha},
that is part of the R\&D activities under
development for the future upgrade of CUORE (the first ton-scale
cryogenic detector in operation looking for {\DBD}\cite{2018PhRvL.120m2501A}).
The CALDER goal is to develop large-area high-sensitivity light
detectors able to measure the very weak Cherenkov light that follows a {\DBD} event
and allow to distinguish it from other backgrounds.
The light is detected by superconducting KIDs~
of few mm$^2$ of active area
deposited on a substrate of several cm$^2$ that acts as a light
absorber and generates phonons that will be absorbed in the superconductor and produce the signal.
The main advantage of using KIDs for this study is that their response can be modeled as a function of measurable parameters of the Mattis-Bardeen theory, so we are 
able to estimate the total energy transformed into quasi-particles, and make a direct comparison with the MC 
results. In addition, the small fractional area covered
by the sensors with respect to the total absorber one
enhances the influence of the phonon reflection/transmission model in the final results.

We apply the G4CMP package to a silicon wafer read by one or several KIDs.
Comparing the simulation results with our data we 
found a notable agreement for the AMM model, and  
we are able to estimate  the transmission coefficients 
at the interfaces Si-Al ({\TSiAl}) and Si-Teflon. 
The results that we present here can be extended to other kind of detectors based on thin Al sensors.
\par The structure of the paper is as follows: 
Section~\ref{sec:mc} presents a brief description of the main physics ingredients included in the MC code 
and the parameters used in our implementation. 
Section~\ref{sec:calder} describes the general aspects of our detectors, 
experimental setup, data analysis, 
and the specific experimental configurations simulated.
The details of the MC simulation are outlined in Section~\ref{sec:g4cmp}, while 
in Section~\ref{sec:results} we compare the MC results with the experimental data 
and make an estimation of the relevant parameters. 
Finally, we present the summary in Section~\ref{sec:discussion}.

\section{Phonon physics} \label{sec:mc}
\indent\indent In this section we describe the basic phonon physics mechanisms implemented in 
the MC simulation, referring to~\cite{Brandt:2014imy,Leman:2011by} for a more complete description. 
Table~\ref{tab:params} reports the numerical parameters used in our 
simulation, whose meaning is given in the following. 
\par
In a phonon-mediated cryogenic detector, particles (optical photons in our case) hitting the
absorber produce optical phonons that decay promptly to the acoustic branch,
producing an athermal population of high energy.
The interaction length of these energetic phonons is very short, so they propagate
quasidiffusively, with numerous changes in direction and polarization mode
as they decay to lower energy states. 
When the phonon energy drops sufficiently, its mean free path
becomes larger than the dimensions of the crystal and it propagates following almost straight lines
at the speed of sound in the material, a state that we call ballistic.
If the dimensions of the
sensor are small compared to the absorber size, as for CALDER detectors, ballistic phonons can
undergo a large number of reflections at the substrate faces before reaching the KID, where they have a 
certain probability {\TSiAl} of being absorbed, or escaping detection (i.e., they are thermalized 
in the substrate or absorbed at the supports or the feedline).
\begin{table}[ht]
\begin{center}
  \begin{tabular}{@{\extracolsep{\fill}} l l l l}
    \hline
    \hline 
    Symbol          & Parameter description & Value                              & Ref \\
    \hline
    d               & density                & 2.33 g/cm$^3$                     & \cite{1985PhRvB..31.2574T} \\
    $C_{11}$        & elastic constant       & 165.6~GPa                         & \cite{Ashcroft76} \\
    $C_{12}$        & elastic constant       & 63.9~GPa                          & \cite{Ashcroft76} \\
    $C_{44}$        & elastic constant       & 79.5~GPa                          & \cite{Ashcroft76} \\
    $\beta$         & 2$^{nd}$ order elastic constant       & -42.9~GPa          & \cite{1985PhRvB..31.2574T} \\
    $\gamma$        & 2$^{nd}$ order elastic constant       & -94.5~GPa          & \cite{1985PhRvB..31.2574T} \\
    $\lambda$       & Lam\'e constant        & 52.4~GPa                          & \cite{1985PhRvB..31.2574T} \\
    $\mu$           & Lam\'e constant        & 68.0~GPa                          & \cite{1985PhRvB..31.2574T} \\
    DOS(L)        & density of states L    &  0.093                              & \cite{1991PhRvB..44.3001T} \\
    DOS(FT)       & density of states FT   & 0.376                               & \cite{1991PhRvB..44.3001T} \\
    DOS(ST)       & density of states ST   & 0.531                               & \cite{1991PhRvB..44.3001T} \\
    $R_A$           & anharmonic decay rate  & 7.41$\times10^{-56}~$s$^4$        & \cite{1993PhRvB..4813502T} \\ 
    $R_I$           & isotopic scattering rate  & 2.43$\times10^{-42}~$s$^3$     & \cite{1993PhRvB..4813502T} \\ 
    ${\nuDebye}$   & Debye frequency           &  15~THz                         & \cite{Ashcroft76} \\
    ${\etapb}$    & pair-breaking efficiency & 0.57                              &\cite{2000PhRvB..6111807K} \\
    $\xi_\textrm{tr}$ & fraction of phonons tracked &  0.02                      & \\
    ${\nmax}$     & maximum number of reflections & 1000                         & \\
    ${\TSiAl}$    & Si-Al transmission coefficient & [0.1 - 1]               & \\
    ${\TSiTef}$   & Si-Teflon transmission coefficient & [0.1 - 1]               & \\
    \hline \hline
  \end{tabular}
  \end{center}
  \caption{Parameters of the G4CMP Monte Carlo simulation. Unless otherwise stated their values are for Si.}
  \label{tab:params}
\end{table}
\par
\begin{figure*}[ht] 
\centering
\includegraphics[width=1\textwidth]{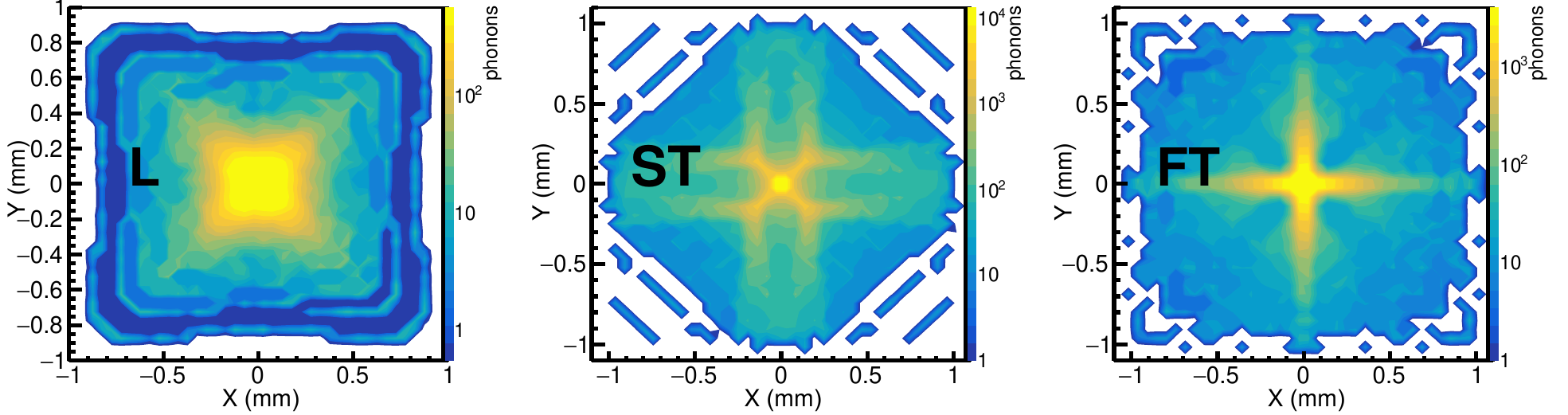}
\caption[]{Simulation of 0.1~THz phonons generated in one small spot of the Si wafer surface and detected in the opposite face. The panels show the flux intensity for every polarization:
L (left), ST (center) and FT (right). The image spans an angle of $\pm$72$^\circ$. Phonon focusing structures (bright colors) are clearly formed, especially for ST and FT modes.
}
\label{fig:caustics}
\end{figure*}

 Phonon tracking in crystalline structures strongly differs from the usual particle propagation
in GEANT code because 
(1) an acoustic phonon can be in three different polarization states, one longitudinal (L) and
two transversal, fast (FT) and slow (ST), with different velocities for every state;
(2) the direction of energy propagation, that occurs along the group
velocity vector, $\nabla_k \omega(\textbf{k})$, where $\omega$ is the phonon frequency,
does not flow in general parallel to the wavevector direction \textbf{k}. This fact,
that depends on the crystal lattice symmetry and physical properties (mainly the elastic constants), 
causes the phonons to
travel in preferred directions along the crystal, a phenomenon known as ``phonon
focusing''~\cite{1969PhRvL..23..416T,1979PhRvL..43.1424N}. 
Silicon has a face centered cubic crystal structure for which we expect
quasi isotropic transport for longitudinal phonons, but 
highly anisotropic one for the transversal modes.
To check that the caustics are correctly generated in our code we perform a 
simulation starting with low energy phonons of around 0.1~THz produced in a small 
spot at the surface of the 
Si wafer. Phonons of this frequency are ballistic in Si, so they propagate along an almost 
unchanged trajectory and polarization state until they reach the opposite face, forming the 
characteristic phonon focusing structures observed in Si by laser-beam experiments~\cite{1985PhRvB..32.2568H} (see Fig.~\ref{fig:caustics}).

\par The phonon propagation in the crystal is mainly governed by two processes:
\begin{enumerate}
\item isotopic scattering: when the substrate is composed of different isotopes, as it is usually the case,
there is a disruption in the propagation path that causes the phonon to scatter off
and change direction with no energy loss. The energy-dependent
rate is modeled as  $R_I\nu^4$,
where $R_I$ depends on the material (see Table~\ref{tab:params}) and  $\nu=\omega/2\pi$ is the phonon frequency.
The single scattering process depends
on the inner product of the polarization vectors of the initial and final phonons, 
but the total expected rate is isotropic. Thus, the isotropic approximation in which the scattered phonon 
has random direction and polarization distribution
according to the density of states (DOS) is pretty accurate after several scatters and much less time consuming;
\item anharmonic decay: due to nonlinear terms in the elastic coupling between adjacent lattice
ions, a phonon spontaneously splits into two (or more) lower frequency ones, 
with a rate that depends on the phonon frequency as $R_A\nu^5$, where $R_A$ 
is a material-dependent constant (see Table~\ref{tab:params}).
A complete treatment of the scattering process is computationally too expensive, so usually
the isotropic approximation, in which only L phonons can decay via L$\rightarrow$L+T and L$\rightarrow$T+T
processes, is adopted.
\end{enumerate}
\par As said before, there is not yet a complete understanding 
of the mechanisms that govern the phonon physics at the interfaces.
Ideally phonons reflect and transmit conserving the energy and the 
component of \textbf{k} parallel to the interface, but polarization conversion can occur, being in general
three reflected waves (trirefringence) and at most two transmitted ones (birefringence)~\cite{2005imph.book.....W}. 
The AMM and DMM models are the most extended but none of them is sufficient to entirely explain 
the experimental data.

\section{Phonon-mediated kinetic inductance detectors} \label{sec:calder}
\indent\indent KIDs operation principle is based on the properties of a superconducting film biased with AC 
(microwave) current.
The inertia of Cooper pairs to momentum change produces an additional inductance, called kinetic inductance (L$_{KI}$), which depends on the
density of Cooper pairs, and that can be measured embedding the 
superconductor in a resonant RLC circuit with resonant frequency 
$\nu_0=1/2\pi\sqrt{LC}$. An energy release larger than twice the 
superconductor gap $\Delta$ (about 200~$\mu$eV for thin Al films) 
breaks Cooper pairs into quasiparticles, modifying
both the residual resistance due to quasiparticles (the only dissipative term in the RLC resonant circuit)
and the inductance due to Cooper pairs and
changing the amplitude and phase of a microwave signal transmitted past the circuit.
Slightly modifying the capacitance of every resonator we can make them 
resonate at close but different frequencies, and in that way many of them can be read with the same line. 
\par
The detector used in this work follows a Lumped Element KID (LEKID) design~\cite{2008JLTP..151..530D}
that uses a separate meander section (inductor) and an interdigital capacitor to form a resonator
coupled (inductively or capacitively) to a Coplanar Waveguide (CPW) for excitation and readout.
They are fabricated
at Istituto di Fotonica e Nanotecnologie of CNR (Rome).
They are pattered by electron beam lithography in a 60~nm Al film deposited by
electron-gun evaporator on a thin ($\sim$300~$\mu$m) high-resistivity Si (100) substrate~\cite{2016NIMPA.824..177C,2016JLTP..184..131C}.
In order to reduce the thermal quasiparticles population
we operate the detector well below the Al critical temperature.
The Si wafer is fixed to a copper holder by small Teflon supports that act as 
thermal link to the heat sink while the holder is anchored to the coldest point of a dilution 
refrigeration, at a base temperature of about 20$~$mK.
\par
KIDs are excited with a fixed-frequency signal 
typically in the few GHz range. After transmission
through the device, the signal $S_{21}$ is amplified by a CITLF4 SiGe
cryogenic low noise amplifier (with noise temperature T$_N\sim$7~K) operated at 4~K and the
rest of the electronics are at room temperature~\cite{Battistelli:2015vha}.
\par 
The signal transmitted through the feedline can be written as a function of the frequency $\nu$ as follows:
\begin{equation} \label{eq:S21}
S_{21}(\nu)={\I} + i{\Q} = 1-\frac{Q/Q_c}{1+2iQ\frac{\nu-\nu_0}{\nu_0}},
\end{equation}
where $S_{21}$ is the forward scattering amplitude in the  standard scattering matrix representation,
{\I} and {\Q} indicate real and imaginary part of $S_{21}$ and
$Q$ is the quality factor of the resonant circuit, which is given by the 
addition in parallel of the 
coupling quality factor $Q_c$ (that account for losses through the coupling) and 
the internal quality factor $Q_i$ (dissipation due to quasiparticles and all other losses), so that 
$Q^{-1} = Q_c^{-1} + Q_i^{-1}$.
When $\nu$ sweeps around the resonance, the signal traces out a circle 
in the ${\I\Q}$ plane of diameter equal to $Q/Q_c$ (see inset of Fig.~\ref{fig:pulse}). We determine the circle center and radius,
taking into account distortions introduced by the power stored in the resonator and 
possible impedance mismatches~\cite{2016JLTP..184..142C},
to translate the {\I}(t) and {\Q}(t) components 
into phase $\delta\phi$(t) and amplitude $\delta a$(t) variations
relative to the center of the resonance loop (calibration).
\par Once the resonance is calibrated we choose the most sensitive frequency 
(or frequencies, in the case of reading several KIDs through the same line) and 
excite the resonators at an adequate power level~\cite{2017ApPhL.110c3504C}.
We run an amplitude threshold trigger algorithm on the continuously acquired signals to capture 
particle passages through the detector, and 
register a window of configurable length around the position of each trigger.
Fig.~\ref{fig:pulse} shows a typical response to a 36~keV energy deposit in the Si substrate. 
The $\delta\phi(t)$ component usually features much better signal-to-noise ratio (SNR)
than $\delta a$(t), 
so in the following we use only this signal.
\subsection{Phonon time constant} \label{sec:pulseTime}

Athermal phonons arrive to the KIDs with a characteristic time distribution that depends on the 
detector material and geometry.
In general it can be modeled by two time constants, accounting for the pulse rise (${\tph}^\textrm{rise}$) and 
decay (${\tph}$), so 
the number of phonons at the KID can be written as:
\begin{equation}\label{eq:phPulse}
N_{ph}(t)=\frac{N_\textrm{ph}}{{\tph}-{\tph}^\textrm{rise}}\left({\expp}^{-t/{\tph}} - {\expp}^{-t/{\tph}^\textrm{rise}}\right).
\end{equation}
When ${\tph}^\textrm{rise}\ll{\tph}$, as in the case of the detectors analyzed in this work, the expression 
\ref{eq:phPulse} can be approximated by a single exponential with constant {\tph}.  \par In order to infer {\tph} from the KID signal, we have to disentangle the contribution
of other temporal constants:
(1) at the KID phonons break Cooper pairs and generate quasiparticles with a probability given by 
the pair-breaking efficiency ${\etapb}$ (see Tab.~\ref{tab:params}), which recombine again into 
Cooper pairs with lifetime ${\tqp}$. The recombination rate depends not only on superconductor properties,
but also on the quasiparticle density, and consequently also on temperature and microwave power ({\power})\cite{2014PhRvL.112d7004D};
(2) the $Q$ factor determines the time constant at which the power dissipation decays as 
{${\tring}=Q/\pi\nu_0$}, hence high-$Q$ resonators are more sensitive but are also slower. 
The temporal evolution of the signal is a convolution of these effects:
\begin{widetext}
\begin{equation}\label{eq:pulse}
\delta \phi(t)=\Phi_\textrm{qp}{\tqp}\left[\frac{{\tqp}{\expp}^{-t/{\tqp}}}{({\tqp}-{\tph})({\tqp}-{\tring})} + \frac{{\tph}{\expp}^{-t/{\tph}}}{({\tph}-{\tqp})({\tph}-{\tring})} + \frac{{\tring}{\expp}^{-t/{\tring}}}{({\tring}-{\tqp})({\tring}-{\tph})}\right],
\end{equation}
\end{widetext}
where $\Phi_\textrm{qp}$ is the pulse integral and its expression is derived in the next section.

\par
As we explain in next section, we excite the substrate by a LED pulse whose duration {\Tex} is of the order of few $\mu$s, 
so the final waveform results from the convolution 
of Eq.~\ref{eq:pulse} with a rectangular function of length {\Tex}.
\par
\begin{figure}[h] 
\centering
\includegraphics[width=0.49\textwidth]{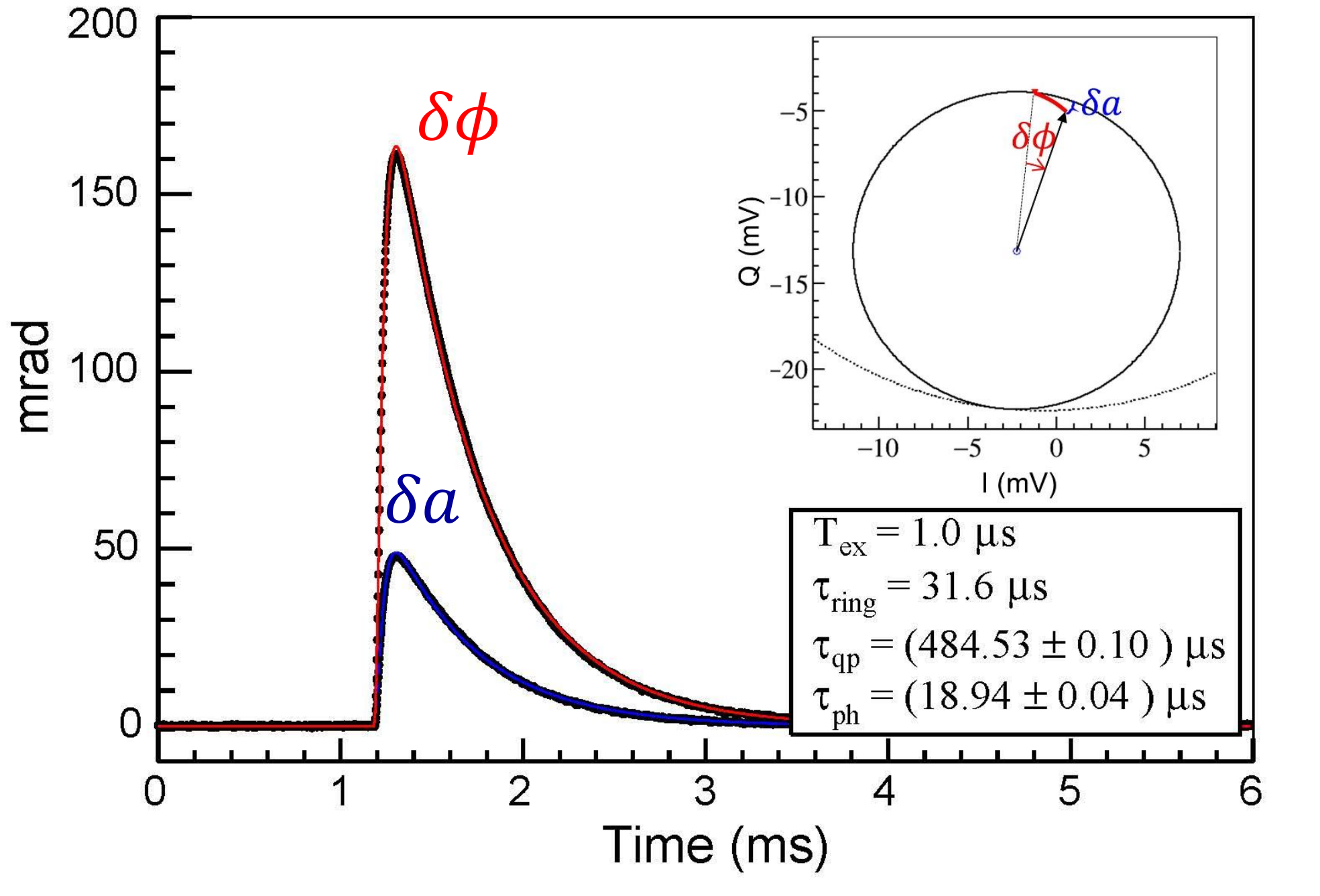}
\caption[]{$\delta\phi$ and $\delta a$ pulse time evolution following an energy deposition of 
36~keV in the Si substrate. The signals are fitted to the pulse shape of Eq.~\ref{eq:pulse}, taking also 
into account {\Tex}=1~$\mu$s (in red 
for $\delta\phi$ and blue for $\delta a$). The $\chi^2$/NDF for the fit in the $\delta\phi$ component is 2.5. The resulting $\delta\phi$ fit parameters 
are shown in the legend. Inset: resonance circle that we calibrate to obtain $\delta\phi$ and $\delta a$ components from
the real and imaginary parts of the $S_{21}$ signal.
}
\label{fig:pulse}
\end{figure}

 For every acquired signal we fit the $\delta\phi$ pulse evolution to the pulse shape described above,
 fixing {\tring} to the value corresponding to the measured $Q$ factor. In this way we obtain 
{\tph} that we compare with the MC results. Superimposed to the pulses of Fig.~\ref{fig:pulse}
we show the results form the fit for the $\delta\phi(t)$ and $\delta a(t)$ signals.

\subsection{Response to energy absorption}
We can relate 
the energy release in the substrate $E$ 
to
the energy absorbed at every resonator ${\Eabs}$ 
through an efficiency factor $\eta$, so that ${\Eabs}=\eta E$.
The efficiency can be factorized as $\eta={\etageom}{\etapb}$, where ${\etageom}$ depends on the
geometry of the detector and the transmission coefficients at the interface, and it is the 
parameter that we shall extract from the MC simulation, and 
${\etapb}$ is the pair-breaking efficiency in Al, which we take as $\sim$0.57~\cite{2000PhRvB..6111807K}.
Now, {\Phiqp} in Eq.~\ref{eq:pulse} represents the overall change in $\delta\phi$ corresponding to an increment 
in the quasi-particle population {\Nqp}={\Eabs}/$\Delta=\eta E/\Delta$, that
can be calculated from the 
Mattis-Bardeen theory in the thin film limit. After some analytical approximations~\cite{Mazin:2005zy} we can write:
\begin{equation} \label{eq:phase}
{\Phiqp}=\frac{\alpha S_2(\nu,{\Tqp})Q}{N_0V\Delta({\Tqp})}\frac{\eta E}{\Delta({\Tqp})},
\end{equation}
where $N_0V$ is the single spin density of states at the Fermi level
(1.72$\times 10^{10}$~eV$^{-1}$~$\mu$m$^{-3}$ for Al~\cite{2003Natur.425..817D}) multiplied by the active volume of the resonator,
$\alpha$ is the fraction of kinetic inductance $L_{KI}/L$,
{\Tqp} is the effective temperature of the quasiparticle system, larger than
the sink temperature due to {\power}, and
$S_2$ is a dimensionless
factor given by the Mattis-Bardeen theory.
The parameters $\Delta$, $\alpha$, $S_2$ and $Q$ are measurable quantities for a given {\power}, therefore 
from the pulse fit we can obtain $\Phi_\textrm{qp}$ and 
determine through Eq.~\ref{eq:phase} 
the efficiency $\eta$ of every pixel 
in order to compare with the MC results.

\subsection{Experimental configurations} \label{sec:exp}
\indent\indent We study two different detector configurations  with different KID characteristics and layout.
\par
\begin{figure}[h] 
\centering
\includegraphics[width=0.49\textwidth]{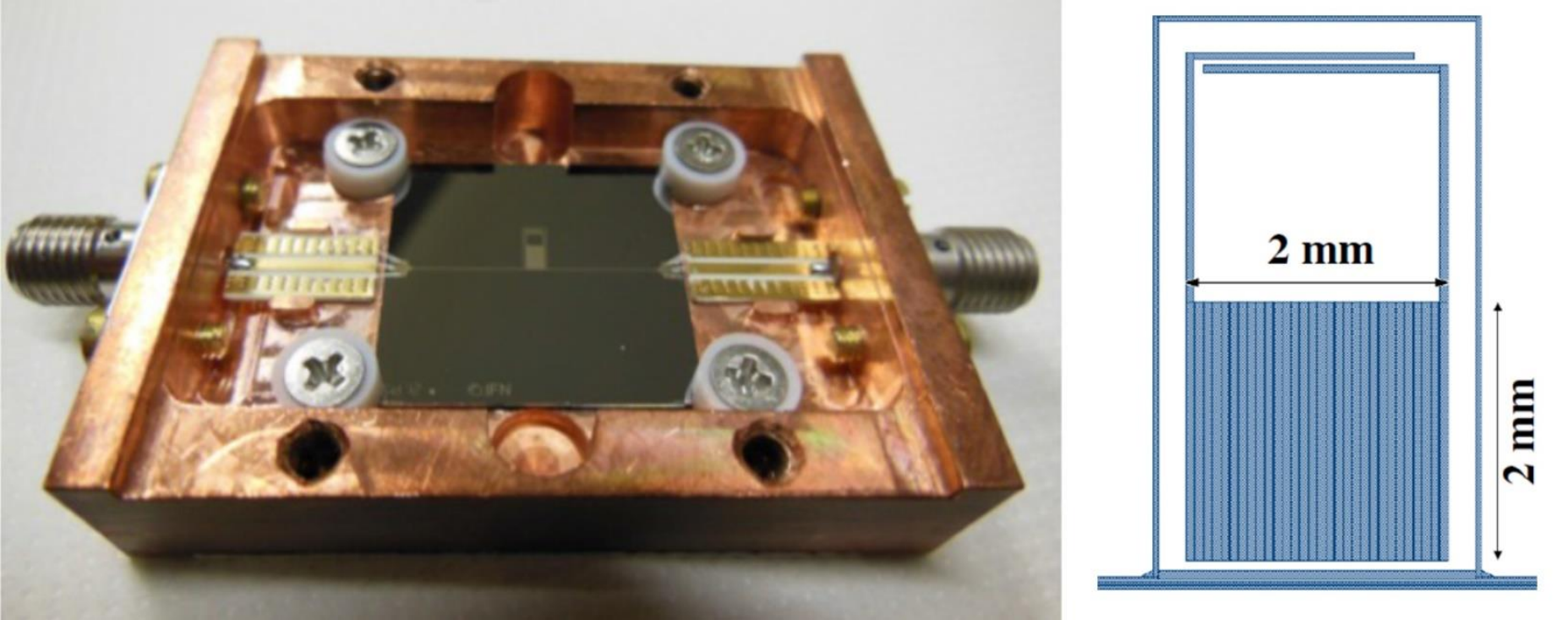}
\caption[]{Left: Picture of the P1 prototype: A 60~nm thick Al KID deposited on a $2\times2~\textrm{cm}^2$
380~$\mu$m thick Si substrate. Four Teflon supports, one at every corner, fix the detector to a copper holder that is anchored
to the cryostat.  Right: geometry of the single pixel: An inductor made of 30 strips of 62.5~$\mu$m$\times$2~mm, 
with gap of 5~$\mu$m between them, and a capacitor composed of two fingers.}
\label{fig:1kid}
\end{figure}

 The first prototype (P1 in the following) consists of a single KID lithographed on 
a $380~\mu m$ thick Si substrate
with a size of 2$\times$2~cm$^2$.
Fig.~\ref{fig:1kid} shows a picture of the detector mounted in the copper holder (left panel) and
a schematic design of the single KID (right panel). 
The inductor section is a meander of 30 strips of 62.5~$\mu$m$\times$2~mm, 
with gap of 5~$\mu$m between them, and the capacitor is composed of only two fingers.
The total active area is 4.0~mm$^2$, excluding the gaps and including the active region that connects the inductor to the capacitor. 
The feedline is a 72~$\mu$m width CPW that cuts across the Si substrate from side to side.
The pixel and feedline are made of 60~nm thick Al. 
Four cylindrical Teflon supports, one at every corner of the substrate, fix the detector to a copper holder that is anchored
to the cryostat. The contact area between Si and Teflon is lower than 3~mm$^2$ at every support.
For detailed results of this prototype, see~\cite{2017ApPhL.110c3504C}.
\par
\begin{figure}[h] 
\centering
\includegraphics[width=0.49\textwidth]{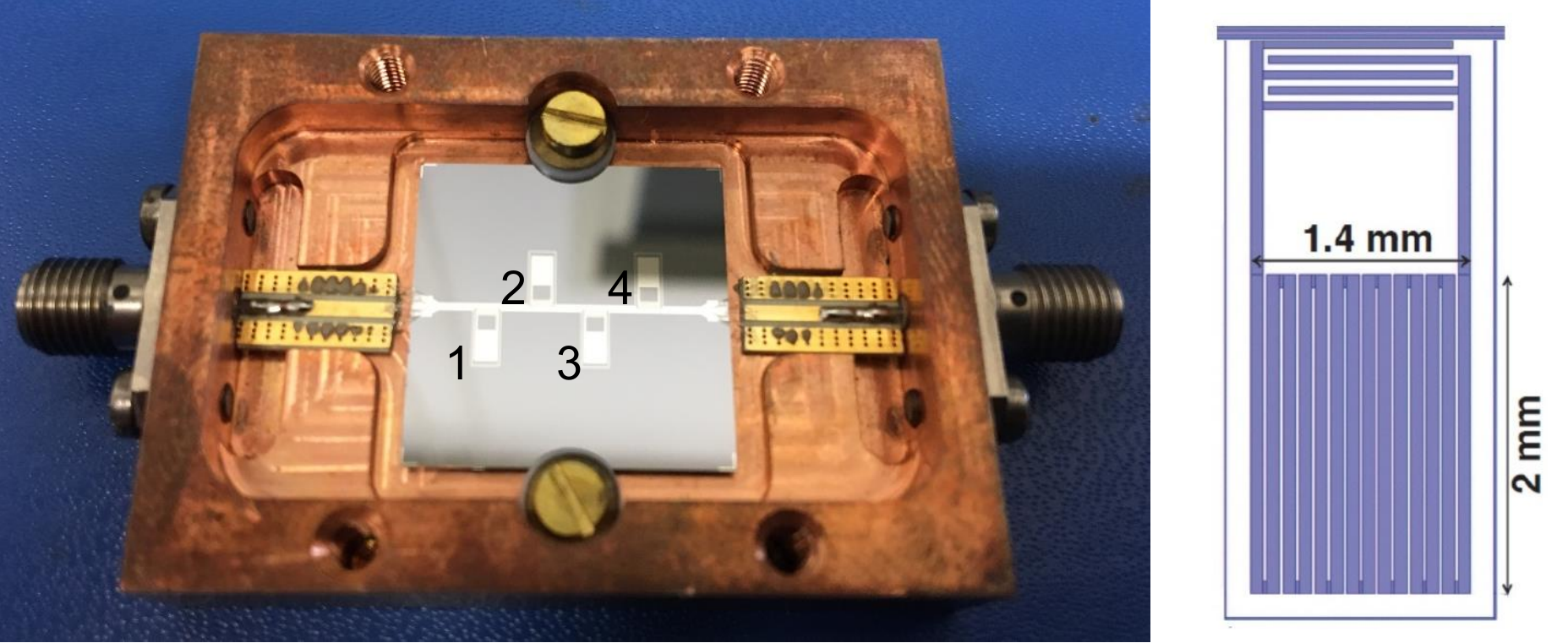}
\caption[]{Left: Picture of the P4 prototype: Four Al KIDs are deposited on a 
300~$\mu$m thick Si substrate with a size of 2$\times$2~cm$^2$.
Two cylindrical Teflon supports with a contact area of around 
3~mm$^2$ each hold the substrate in the copper structure.
Right: geometry of the single pixel (60~nm thick Al film): an inductive meander with 14 connected strips
of 80~$\mu$m$\times$2~mm and capacitor made of 5 interdigitated fingers of 1.4~mm$\times$50~$\mu$m. 
The active area of the single pixel is $1.15\times2$mm$^2$.}
\label{fig:4kid}
\end{figure}

In the second prototype, that we label as P4 (see Fig.~\ref{fig:4kid}), the wafer is 
$375~\mu m$ thick and there are four Al KIDs 
with an inductive meander made of 14 connected strips
of 80~$\mu$m$\times$2~mm closed by a capacitor
made of 5 interdigitated fingers of 1.4~mm$\times$50~$\mu$m. The active area of the single pixel is $1.15\times2$mm$^2$.
The feedline is a 420~$\mu$m width and 60~nm thick CPW.
\par Compared to P1, P4 has smaller contact area between Si and Teflon, as it is held by only two supports at opposite edges in the middle 
of the substrate. The contact area at every support is about 3~mm$^2$, so the total 
interface Si-Teflon is halved with respect to P1. In turn, the feedline is $\sim$6 times wider.
\par
\begin{table*}[ht]
\begin{center}
  \begin{tabular}{@{\extracolsep{\fill}} l l p{1.8cm} p{1.5cm} p{3cm} l l l l l l l l l}
    \hline
    \hline
    Prototype & coupling & T$_c$ & $\Delta_0$ & $\alpha$ & $P_{\mu\nu}$ & KID & $Q$ & $Q_i$ & $Q_c$ & $\tau_\textrm{ring}$ & \multicolumn{3}{c}{Source}\\
              &          & [K] & [$\mu$V] & [\%] & [dBm] & & [k] & [k] & [k] & [$\mu$s] & pos [mm] & $\phi$ [mm] &  {\Tex} [$\mu$s] \\
    \hline
    P1    & inductive & 1.180$\pm$0.020 & 179$\pm$3 & 2.54 $\pm$0.9$_{stat}\pm$0.26$_{syst}$ & -76.8 & 1 & 149 & 2301 & 159 & 18.2 & (0,-6) & 4.66 & 10 \\
\hline
    \multirow{4}{*}{P4} & \multirow{4}{*}{capacitive} & \multirow{4}{*}{1.300$\pm$0.025}  & \multirow{4}{*}{197$\pm$4}  & \multirow{4}{*}{2.14$\pm$0.04$_{stat}\pm$0.27$_{syst}$}  & \multirow{4}{*}{-79.1} & 1 & 18.6 & 69.7 & 25.4 & 2.23 & \multirow{4}{*}{ (0,0) } & \multirow{4}{*}{4.66} & \multirow{4}{*}{1} \\
& & & & & & 2 & 38.4 & 99.6 & 62.4 & 4.59 \\
& & & & & & 3 & 138 & 899 & 162 & 16.4 \\
& & & & & & 4 & 266 & 407 & 772 & 31.6 \\
    \hline \hline
  \end{tabular}
  \end{center}
  \caption{Experimental relevant parameters of the simulated experiments with P1 and P4 prototypes. See text for details.}
  \label{tab:Q}
\end{table*}
\par We operate both prototypes as described at the beginning of this section.
The first step is to select the excitation power {\power}. 
High powers feature in general a better SNR, as the noise in our setup is dominated by the amplifier and 
goes with $1/\sqrt{\power}$, but as we raise {\power} the resonances 
show an increasing distortion and the relationship of Eq.~\ref{eq:phase} is not longer valid~\cite{2017ApPhL.110c3504C}.
Therefore we perform a power scan and select the largest {\power} before distortion.
\par For every prototype we measure the parameters that enter in Eq.~\ref{eq:phase} and report their values 
at the selected {\power} in Table~\ref{tab:Q}. 
The $Q$, $Q_i$ and $Q_c$ factors are computed by fitting the resonance circle as described in \cite{2016JLTP..184..142C}.
We determine the critical temperature $T_c$
(that for thin films depends on thickness and other parameters, as the 
quality of the deposition) during the cooling-down and infer $\Delta_0$ from BCS theory.
Then, we compute $\alpha$ from the resonant frequency shift as we increase the thermal quasiparticle density by 
increasing the base temperature of the system. We fit the resulting curve to the
Mattis-Bardeen theory prediction~\cite{2006NIMPA.559..585G}, keeping $\Delta_0$ fixed in the fit.
For the P4 prototype we average the results of the four resonators.
\par The detectors are illuminated on the back of the substrate by an optical fiber
coupled to a fast warm LED ($\lambda=$400~nm).
The LED equivalent energy is calibrated
with a photomultiplier, and the calibration is checked at
very low intensity by photon counting Poisson statistics~\cite{2018SuScT..31g5002C}.
In Table~\ref{tab:Q} we report also the source position with respect to the center of the substrate,
the diameter of the illuminated spot ($\phi$) and the optical pulse duration.
\par We take $\mathcal{O}(2000)$ LED pulses for every configuration. 
In order to improve the SNR we apply 
a software low-pass filter with 100~kHz cut-off whose effect is included in the pulse fitting.
Finally, we average the pulses and 
perform a fit as described in Sec.~\ref{sec:pulseTime} to 
obtain {\tph} and $\eta$.
We report the results for each KID in Table~\ref{tab:data}. 
\begin{table}[ht]
\begin{center}
  \begin{tabular}{@{\extracolsep{\fill}} p{2cm} p{1cm} p{1.5cm} p{3.5cm}}
    \hline
    \hline
    Prototype & KID & ${\eta}$ [\%] & ${\tph}$ [$\mu$s]  \\
    \hline
                      P1  & 1 & 13.3$\pm$1.1   & $25.4\pm0.1_{stat}\pm0.2_{syst}$ \\
    \hline
    \multirow{4}{*}{P4}   & 1 & 2.9$\pm$0.3   & $16.8\pm1.4_{stat}\pm2.3_{syst}$ \\
                          & 2 & 6.7$\pm$0.7 & $8.64\pm0.14_{stat}\pm0.84_{syst}$ \\
                          & 3 & 6.2$\pm$0.7 & $9.09\pm0.05_{stat}\pm0.56_{syst}$ \\
                          & 4 & 2.7$\pm$0.4 & $15.5\pm0.5_{stat}\pm2.8_{syst}$   \\
    \hline \hline
  \end{tabular}
  \end{center}
  \caption{Experimental results of the P1 and P4 prototypes. For every KID we report 
the efficiency ${\eta}$ and the characteristic phonon arrival time ${\tph}$.
}
  \label{tab:data}
\end{table}
The error in $\eta$ is dominated by the systematic error in $\Delta_0$ and $\alpha$.
For {\tph}, in addition to the statistical error of the fit, we estimate a systematic one by starting from different 
sets of initial fit parameters and by taking pulses with different {\Tex} ranging from 1 to 10~$\mu$s.
The $\chi^2$/NDF of the fits range between 1 and 3.5 for all the KIDs except for KID3, for which we obtain 
values between 4 and 6.8.
In P4 prototype there is a very small ($\sim$200~$\mu$m) rightwards shift 
of the kids layout with respect to the center of the substrate.
It is not appreciable in Fig.~\ref{fig:4kid}, but it is responsible for the slight ($\sim$7\%) 
larger efficiency of KIDs 1 and 2 with respect to KIDs 3 and 4, as they are slightly closer to the source. 
The simulation also includes the shift, so we expect to observe this small effect in the MC results as well.

\section{G4CMP MC implementation}\label{sec:g4cmp}
\par The G4CMP package simulates: (1) the generation of acoustic phonons and electron-hole pairs in a material 
after an energy deposition; (2) their propagation in the media, anisotropic according to the material elastic constants for 
phonons and driven by an electric field for the charge; (3) the two principal phonon scattering processes described 
in Sec.~\ref{sec:mc} with isotropic approximation; and (4) a simplified implementation of the reflection and transmission mechanisms at interfaces, 
in which the multirefringence is not considered: the phonon is transmitted through the boundary with a probability given by the transmission coefficient, or it is reflected back, following a specular reflection for the AMM model or a Lambertian one for DMM. So, in the current implementation no mode conversion occurs.
\par In our simulation, as no electric field is applied 
to the detector, the charge is not taken into account and
all the energy of the interaction goes to the phonon channel.
Following a photon absorption in the Si substrate, acoustic phonons 
are generated isotropically along the incident particle track. 
The energy distribution of the primordial phonons is unknown, nevertheless their effects are wiped out after the
quasidiffusion stage, so we take the Debye energy ($\sim$62~meV in Si) as starting point
and select the polarization L, FT or ST randomly according to the DOS in the material.
The history of every phonon is followed recording its polarization, $\omega$ and \textbf{k} until
one of the following conditions is verified:
(1) it is absorbed in Al (KIDs or feedline) or Teflon;
(2) its energy drops below $2\Delta$; (3) a predefined number of reflections {\nmax} is
reached.  
\par
\begin{figure}[h] 
\centering
\includegraphics[width=0.49\textwidth]{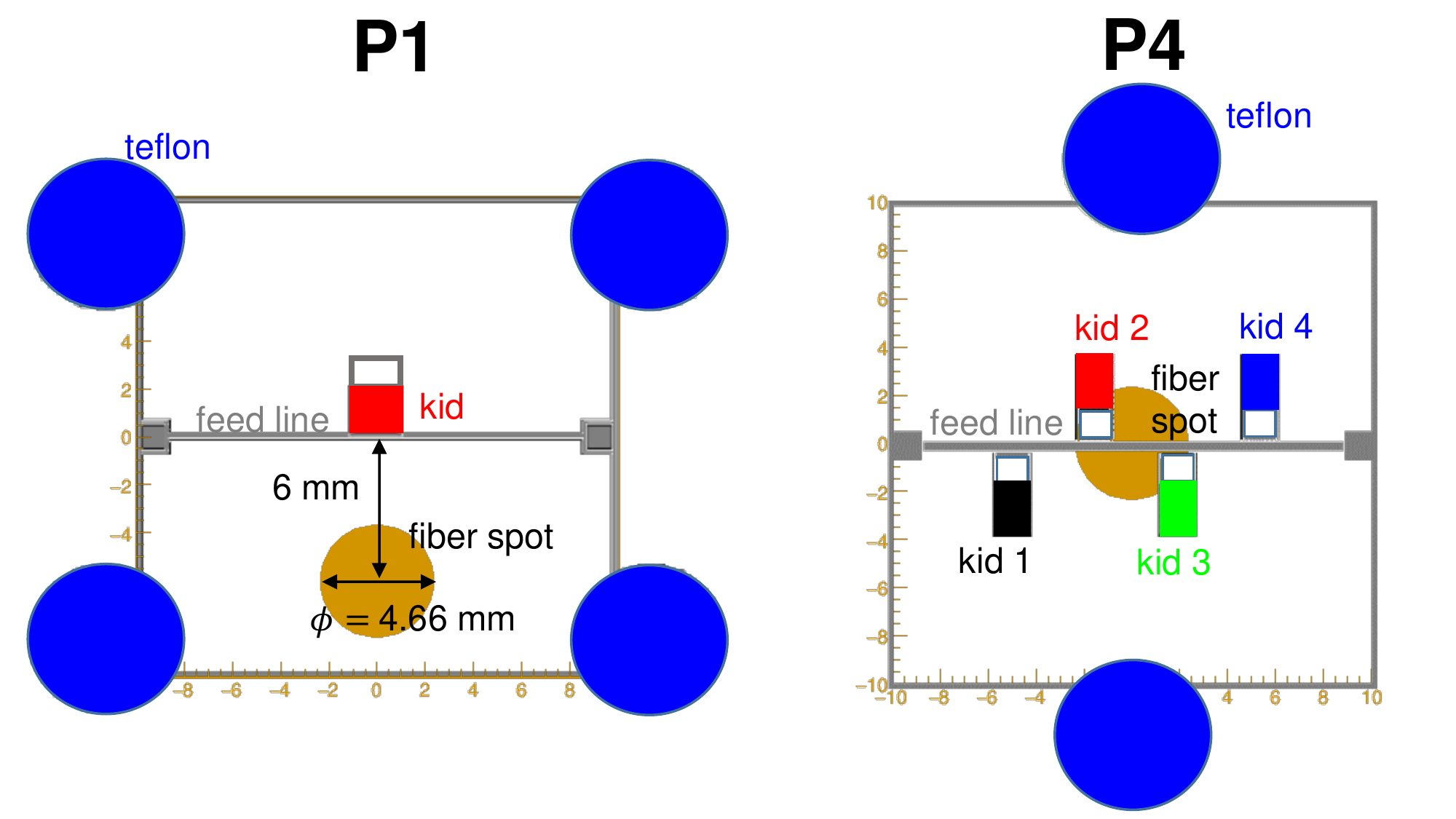}
\caption[]{A sketch of the main components included on the MC simulation of the prototype P1 (left)
and P4 (right). In both cases the fiber spot (brown) has a diameter of 4.66~mm and fires on the opposite side of the KIDs.}
\label{fig:simuGeom}
\end{figure}

 We simulate a simplified geometry of the detector with four main components:
Si wafer, Teflon supports, the feedline and the KIDs, both made of Al (see Fig.~\ref{fig:simuGeom}).
For the sake of keeping the simulation computing time at a reasonable level, 
only a certain fraction of the phonons $\xi_\textrm{tr}$ are tracked (see Tab.~\ref{tab:params})
and the final results are scaled with this value.
In order to determine the effect of the reflection model and 
transmission coefficients
we generate a batch of simulations spanning {\TSiAl} and {\TSiTef} from 0.1 to 1, for both models.
It is worth mentioning that the code does not implement phonon propagation in Al, so a phonon absorbed in the KIDs generates 
a signal with probability {\etapb} or is killed. Hence, (1-{\TSiAl}) include the probability of a phonon to 
enter the Al and to be reflected back to the Si substrate.
\par 
A single simulation event starts with the generation of about 10$^4$ 
optical photons ($\lambda$=400~nm), uniformly distributed at the 4.66~mm diameter fiber spot, that are stopped in the 
first micron of the Si substrate at the opposite face to the KIDs.
The spot is centered in the middle of the substrate in P4 simulation, while for P1 it is shifted 6~mm far from the KID in vertical direction, and the photons are distributed in time according to a square pulse of duration {\Tex} (see Table~\ref{tab:Q}).
For every configuration we generate between 20 and 50 single events.
The outputs of the simulation are the time, energy, position and polarization of every phonon absorbed
in the Teflon, the feedline or the KIDs. 

\section{Results and discussion}
\label{sec:results}
\indent\indent For each fiber event in the wafer 
we construct the phonon pulse evolution for every time 
and integrate it to obtain the total energy absorbed in the simulation at every KID, that we denote as $\Delta$E$_{ph}$.
Then, we scale with the number of tracked phonons $\xi_\textrm{tr}$ and the pair-breaking efficiency {\etapb}
to calculate the absorbed energy, and divide by E to obtain the efficiency in a single KID as
\begin{equation}
\eta=\frac{1}{E}\frac{\etapb}{\xi_\textrm{tr}}\Delta E_{ph}.
\end{equation}

\begin{figure}[h] 
\centering
\includegraphics[width=0.49\textwidth]{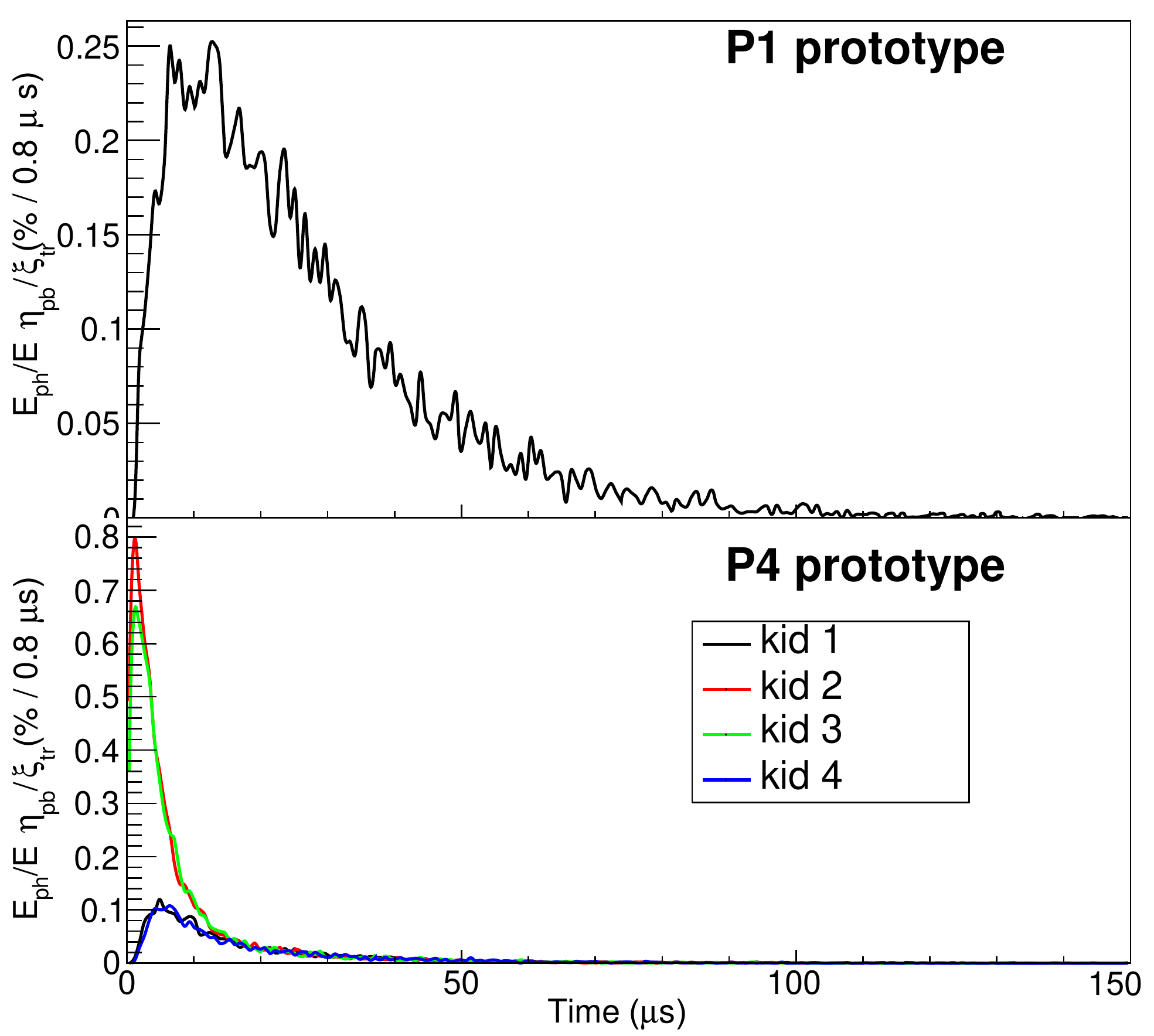}
\caption[]{Phonon distribution at the KIDs for P1 (upper panel) and P4 (bottom panel), corresponding to AMM model, {\TSiAl}=0.36 and {\TSiTef}=0.4.}
\label{fig:pulsesPh}
\end{figure}

Fig~\ref{fig:pulsesPh} displays one of such events for AMM model, {\TSiAl}=0.36 and {\TSiTef}=0.4 for both prototypes.
The simulation does not include resonator-related time constants ({\tring}, {\tqp}), so the pulse shape is described by 
Eq.~\ref{eq:phPulse}.
The rise time of the phonon pulses is around one order of magnitude 
smaller than the decay time, so we consider only one time constant {\tph}
calculated as {\tphSim}/2.2, where {\tphSim} is the 
90$^{th}$ minus the 10$^{th}$ percentile of the phonon distribution.
\par We observe no substantial variations in arrival time among the three polarizations, despite their different 
velocity ($\sim$9000~m/s longitudinal, $\sim$5400~m/s for the transversal modes) since modes are highly 
mixed as a consequence of the scattering processes. For example, for the P1 pulse in Fig.~\ref{fig:pulsesPh} 
we obtain {\tph}=(21.3,~21.3,~21.0)~$\mu$s for (L,~FT,~ST) components separately and {\tph}=21.2~$\mu$s for the three modes together. 
\par The choice for the {\nmax} parameter is not of great importance in the final results: 
for the configurations with low values of the transmission coefficients 
({\TSiAl}$\sim$0.2, {\TSiTef}$\sim$0.1),
only 1-3\% (0.1-0.4\%) phonons undergo more than 200 (500) reflections. For values of {\TSiAl} and {\TSiTef} around 0.4, 
the percentages are 0.5-1\% (0.05-0.1\%).
\par
\begin{figure}[h] 
\centering
\includegraphics[width=0.49\textwidth]{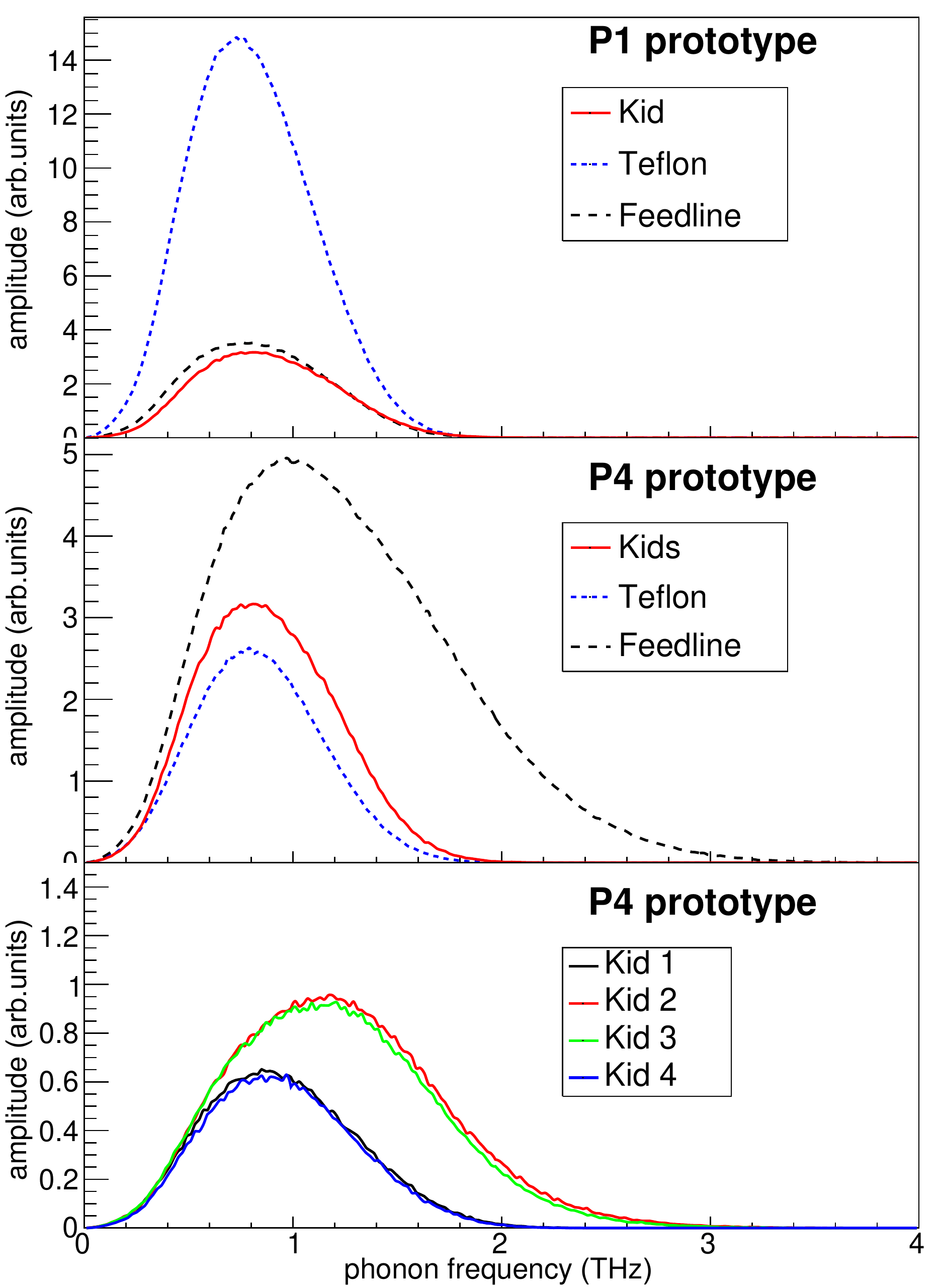}
\caption[]{Frequency distribution of the phonons absorbed in the different materials (Teflon, feedline, KIDs) for P1 (upper panel)
 and P4 (middle panel). In the bottom panel the P4 distribution at every KID is plotted separately.}
\label{fig:freqDistrib}
\end{figure}

 We also study the amount of phonon absorption at every material as a function of phonon frequency
and show the results in Fig.~\ref{fig:freqDistrib} for the same configuration as Fig.~\ref{fig:pulsesPh}. 
The geometric differences among the two prototypes described in Sec.~\ref{sec:exp} 
(more Teflon in P1, $\sim$6 times wider feedline in P4) are clearly reflected in the simulation: 
while for P1 most of the phonons are absorbed in Teflon (about 60\% of the total), in P4 
the element that is taking the major part is the feedline ($\sim$55\%), followed by the KIDs ($\sim$28\%) and then the Teflon ($\sim17$\%).
The maximum of the distributions are at phonon frequencies between 0.7 and 0.9~THz and they are slightly asymmetric, with positive skewness. 
When the origin of the phonon pulse is close to the absorbing element, as for the feedline and KIDs 2 and 3 in P4,
 the asymmetry is more pronounced with a longer tail to higher frequencies.
\par
\begin{figure}[h] 
\centering
\includegraphics[width=0.49\textwidth]{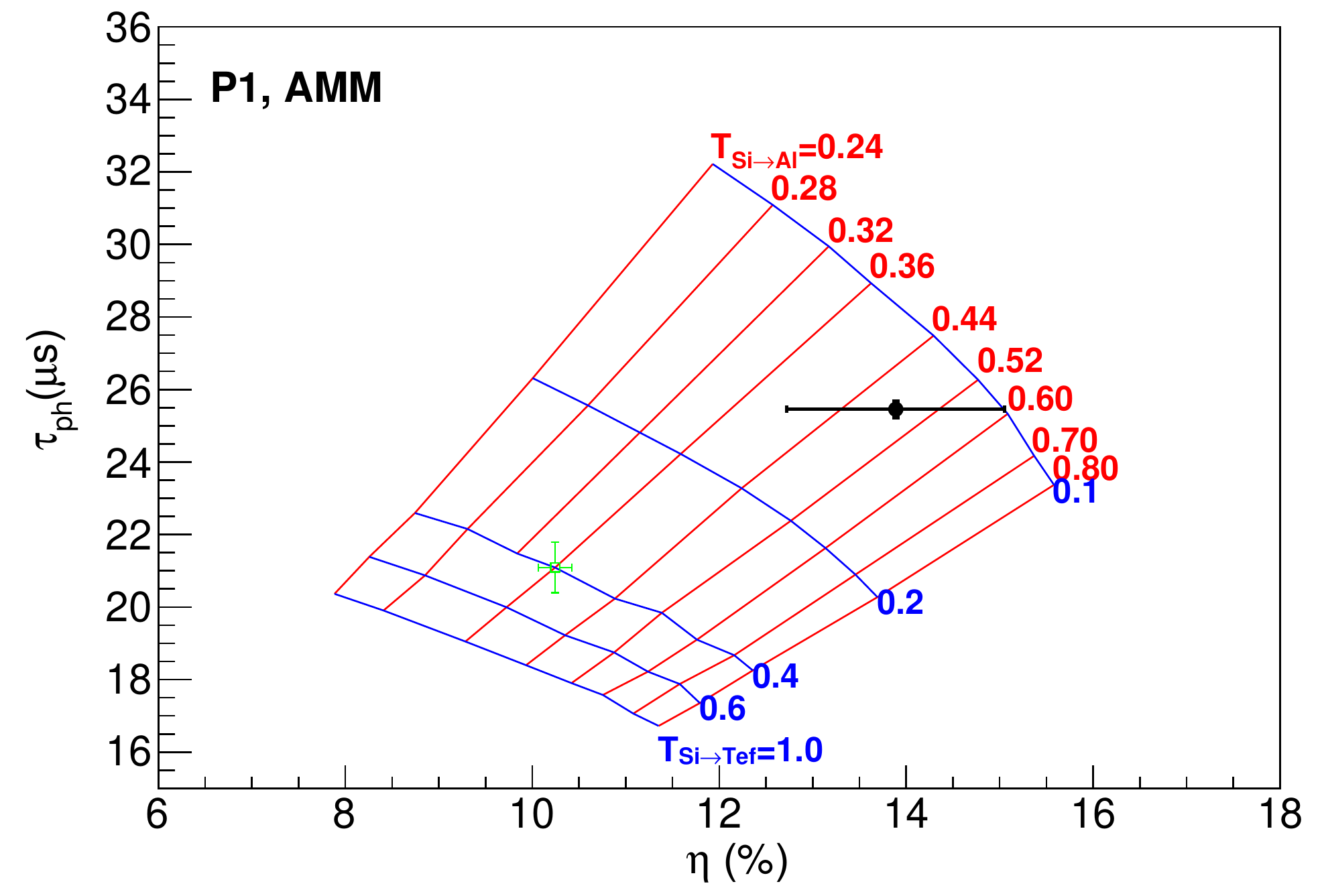}
\includegraphics[width=0.49\textwidth]{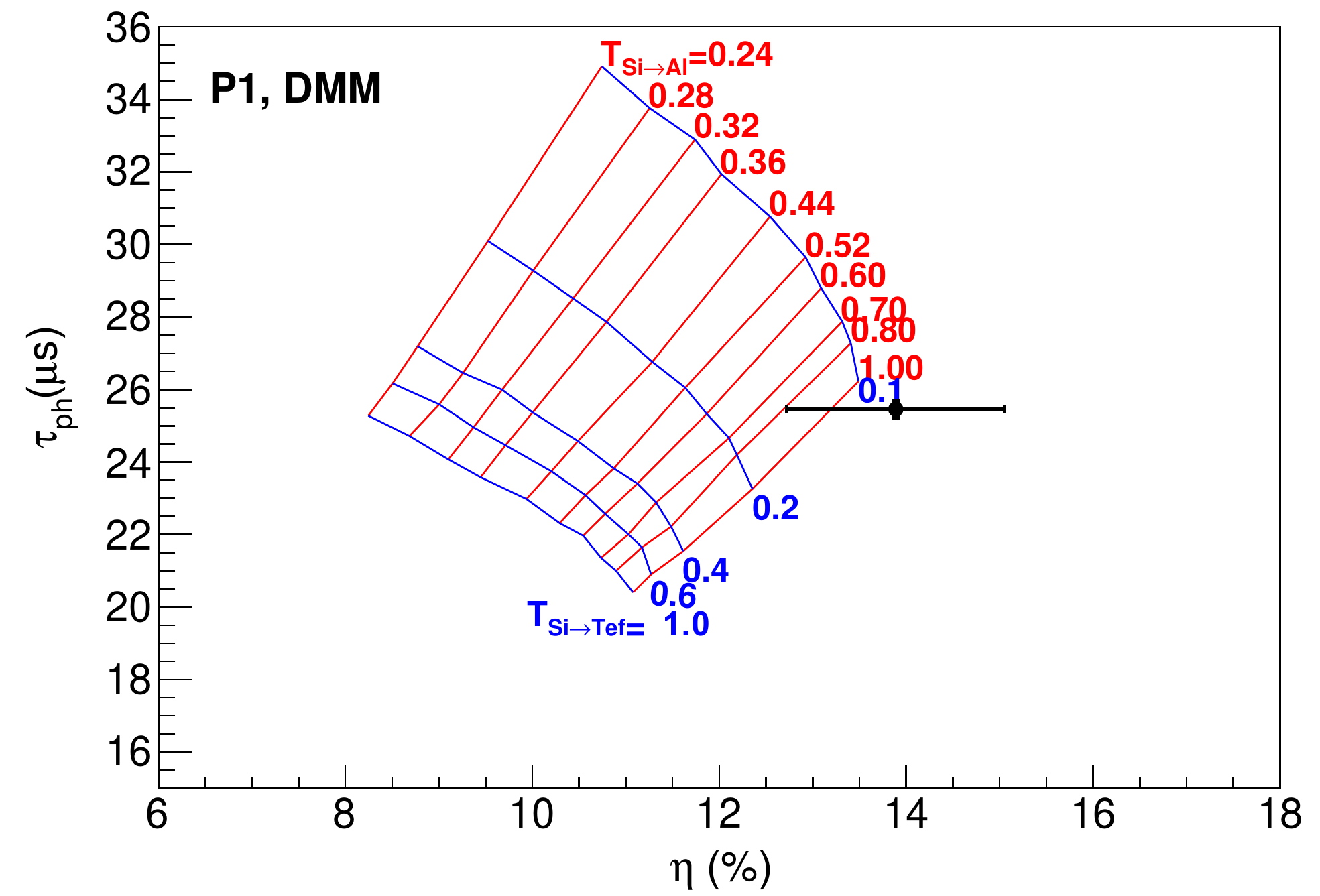}
\caption[]{Comparison of the MC results with experimental data for P1 prototype and AMM model (upper panel)
or DMM model (bottom panel).
The red (blue) lines  correspond to simulations with constant values of the 
{\TSiAl} ({\TSiTef}) coefficient, while the points are taken from Table~\ref{tab:data}. The green error
bars represent the systematic uncertainty associated to the MC parameters.}.
\label{fig:1kidResults}
\end{figure}

\begin{figure*}[t] 
\centering
\includegraphics[width=1\textwidth]{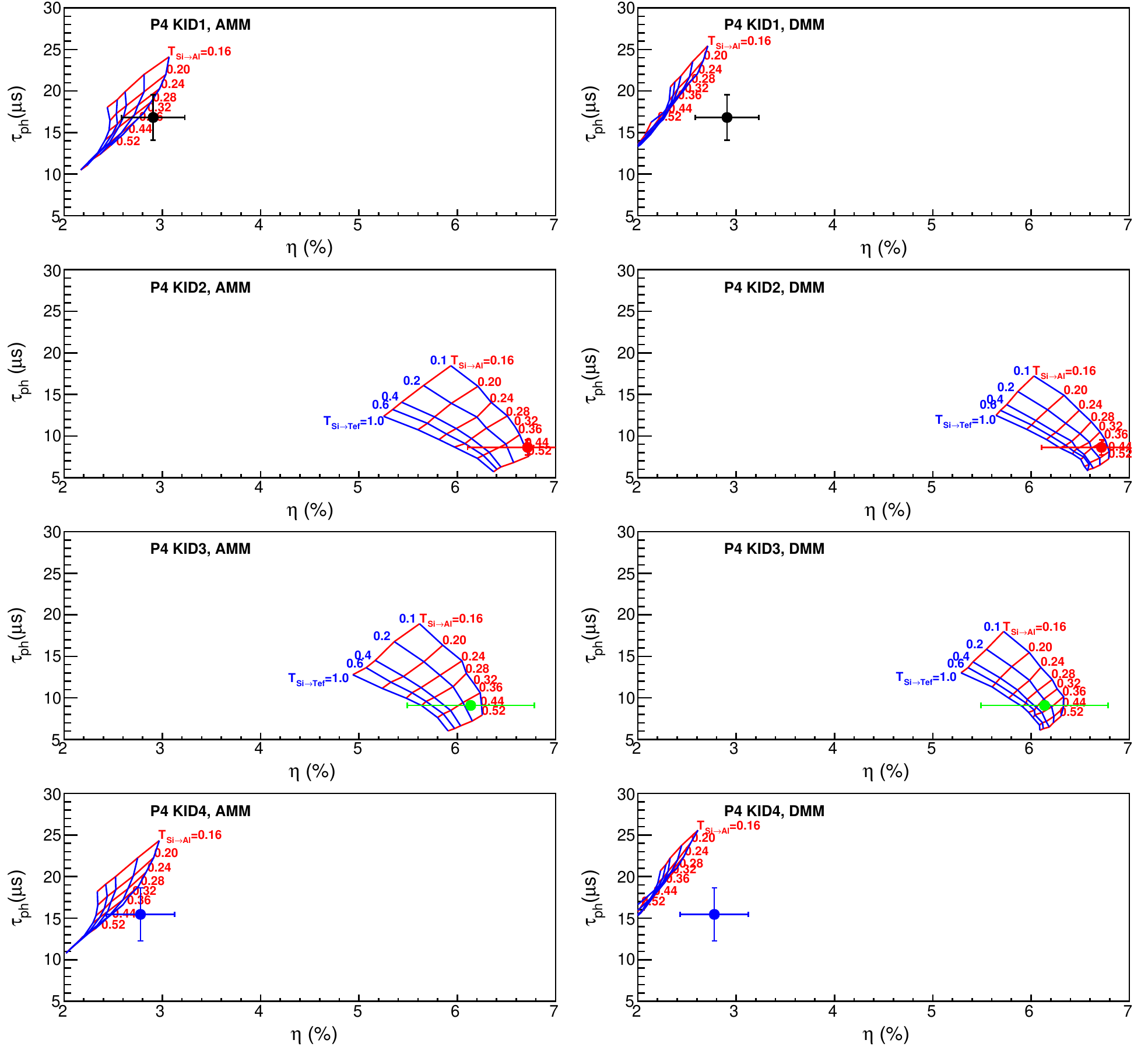}
\caption[]{Same as Fig.~\ref{fig:1kidResults} for the four KIDs of P4.}
\label{fig:4kidResults}
\end{figure*}

 Finally, in Figs.~\ref{fig:1kidResults} and \ref{fig:4kidResults} we compare 
the MC results with the experimental data.
The red (blue) lines  correspond to simulations with constant values of the 
{\TSiAl} ({\TSiTef}) coefficient, while the points are taken from Table~\ref{tab:data}.
In order to estimate a systematic error associated to the simulation 
we have identified the most sensitive parameters of the model to be the decay constants 
$R_A$ and $R_I$ and the elastic constants $C_{11}$, $C_{12}$ and $C_{44}$. We have considered a variation of $\pm$5\% for the elastic 
constants~\cite{1964JAP....35.3312M, 1985PhRvB..32.3792N, CHEN19921} and 
$\pm$20\% for the decay constants and 
calculated the variation in {\tph} and $\eta$ for some simulated configurations. The result for AMM model,
{\TSiAl}=0.36, {\TSiTef}=0.4 is $\pm$3\% in {\tph} and $\pm$2\% in $\eta$ (green lines in Fig.~\ref{fig:1kidResults}).
Similar results are obtained for other configurations.  
As regards the fraction of tracked phonons $\xi_\textrm{tr}$, increasing it from 2\% 
up to 20\% produces an error below 0.2\%.

For P1, with one single KID, phonon pulses are faster and more energetic  
for larger {\TSiAl} values. When we increase {\TSiTef} instead, they are also 
faster, but less energy is collected, as phonons are lost in Teflon.
This rule no longer holds true when more than one KID is competing for the same 
energy deposition, as it is the case of P4: the sensors far from the source 
(KID1 and KID4 in Fig.~\ref{fig:4kidResults}) reverse behavior, and the collected 
energy is lower for larger values of {\TSiAl} because it is being more quickly absorbed in the near KIDs
and the feedline. 
The small shift in the KIDs position towards the right side of the wafer in P4 is 
also noticeable in the simulation and results in larger energy depositions in 
KID1 and KID2 compared to those of KID3 and KID4.

\par In general, simulations with DMM model produce slower and less energetic 
phonon pulses than those with AMM, except when KIDs are very close to the phonon
source, as it is the case of KIDs 2 and 3 in P4.
An explanation for this behaviour can be found in the very different 
propagation patterns that phonons follow once they enter the ballistic regime for 
specular or diffusive reflection. In our geometry we observe 
a much larger density of phonon tracks in the 
central part of the wafer for the AMM simulation rather than for DMM.
The origin of this different distribution
could be, as pointed out by some authors \cite{1984PhRvL..52.2156N},
that phonon caustics survive up to some degree with the specular reflection 
while a more homogeneous distribution of phonons is expected for a Lambertian reflection 
For our geometry, the larger concentration of phonons in the central part 
of the wafer results in a more effective energy collection at the KIDs than in Teflon.
\par
We obtain a consistent picture between data and simulation for both prototypes 
for AMM model, while our experimental data cannot be modelled 
considering only diffuse reflection, unless extreme values of the 
transmission coefficients are introduced. 
In the case of P1 (the most sensitive probe for the reflection model, 
as the energy deposition is far away from the KID), 
for the same transmission coefficients, the DMM model produce phonon pulses 
between 2 and 4 $\mu$s slower. 
This corresponds to between 10 and 20 times our experimental uncertainty. 
Our simulation with DMM model is not able to produce so fast an energetic 
phonon pulses as the ones we have measured for this setup, 
unless a transmission coefficient of almost 1 is considered for {\TSiAl}.
\par
The range of values of {\TSiAl} that 
best describe the experimental data is [0.30-0.55].
In the case of {\TSiTef}, P1 data points to 
the region [0.1-0.15], nevertheless 
the P4 simulations do not impose a large constraint, 
as in general the whole {\TSiTef} range
agrees with the experimental point at 1$\sigma$ error as a result of the reduced Si-Teflon 
interface.
At a closer look, the AMM P4 simulation could be affected by a systematic bias:
in the MC less energy is collected at the KIDs far from the source (KID1 and KID4) 
with respect to the measurement.
This distance-dependent bias could suggest a deficiency of the model 
that appears when the number of phonon reflections is large.
This could be due to the simplification of the reflection mechanisms in the simulation, 
that currently do not include mode conversion, or other phenomena non considered in the 
present implementation.
For example, a slight dependence of the transmission coefficients with phonon frequency would 
result in distinct absorption for far and near KIDs, as the phonon frequency 
distribution is different (see Fig.~\ref{fig:freqDistrib}).
A larger substrate and/or a different KID layout would be necessary to test this conjecture.
\par It is worth noting  that 
in general we expect the phonon transmission coefficient to be dependent on the thickness for thin films. 
The experimental data presented here correspond to an Al thickness 
of 60~nm, and so does the {\TSiAl} transmission coefficient that we have determined. 
Future measurements with 
different films will allow us to study this dependency.

\section{conclusion}
\label{sec:discussion}
We have implemented a phonon MC simulation based on the G4CMP extension of the GEANT4 code 
and applied it to model phonon-mediated cryogenic detectors with thin Si absorbers and Al KID readout, clamped by 
Teflon supports to a dilution unit at about 20~mK.
We have performed two different experiments with different geometries and KID layouts and we have compared 
the results with those of the MC simulations, considering two different reflection mechanisms at the interfaces
(a specular reflection based on the AMM model and a diffuse one for the DMM model) and transmission coefficients spanning 
form [0.1-1] for the Si-Teflon and Si-Al interfaces.
We found a good agreement for transmission coefficients Si-Al in the range [0.3-0.55]  and Si-Tef in the range [0.1-0.15] for AMM model,
while the simulation with diffuse reflection based on the DMM model
does not provide a realistic description of our data.
The Si-Al result is valid for an Al film with a thickness of 60~nm.
We observe also a hint of a systematic bias in our simulation when the number of 
phonon reflections is large: simulated phonon pulses are 
less energetic than data. In the future we will further investigate this 
issue with larger detectors.
The results that we have presented are applicable to other cryogenic detectors with thin Al sensors.

\section*{Acknowledgments}
This work was supported by the European Research 
Council (FP7/2007-2013) under Contract No. CALDER No. 
335359 and by the Italian Ministry of Research under the 
FIRB Contract No. RBFR1269SL.
The authors thanks the personnel of INFN Sezione di Roma for the technical support,
in particular M. Iannone.


\bibliographystyle{apsrev}



\end{document}